\documentstyle[12pt,psfig,twoside]{article}
\newcounter{muni}
 
\newenvironment{remunerate}{\begin{list} 
{
{\rm \arabic{muni}.}}{\usecounter{muni}
\setlength{\leftmargin}{0pt}\setlength{\itemindent}{38pt}}}{\end{list}
} 
\begin{document}
\hbadness=10000
\pagenumbering{arabic}
\pagestyle{myheadings}
\markboth{J. Letessier, J. Rafelski and A. Tounsi}{Strange Particle
Freeze-Out}
\title{STRANGE PARTICLE FREEZE-OUT}
\author{$\ $\\
\bf Jean  Letessier$^1$, Johann Rafelski$^{1,2}$ {\rm and} Ahmed
Tounsi$^1$\\ $\ $\\
$^1$ Laboratoire de Physique Th\'eorique et Hautes Energies\thanks{\em
Unit\'e  associ\'ee au CNRS UA 280, \rm\newline
\hspace*{0.5cm}
Postal address: LPTHE~Universit\'e PARIS 7, Tour 24, 5\`e
\'et., 2 Place Jussieu, F-75251 CEDEX 05.}~, Paris
\\ 
$^2$ Department of Physics, University of Arizona, Tucson, AZ 85721\\
}
\date{}   
\maketitle 
\vspace{-9cm}\noindent {PAR/LPTHE/93--54}  \hfill{November
1993}
\centerline{\ \ \ \ Published in Phys. Lett. B 321  (1994) 394.}
\vspace{7cm} 
\begin{abstract}
{\noindent
We reconsider thermal conditions of the central fireball presumed to
be the source of abundantly produced strange (anti-)baryons in S $\to$
W collisions at 200 GeV A. We show that it is possible to completely
fix the freeze-out temperature of strange particles in terms of the 
central rapidity kaon to Lambda particle abundance ratio at fixed,
high transverse mass using a non-equilibrium hadronization model and
the measured quark fugacities.
}\end{abstract}  \vspace*{0.5cm}
Kinetic strange particle production models \cite{Raf82} imply that
abundant strangeness is suggestive of the quark-gluon plasma (QGP).
Even more specific information about the nature of the dense matter
formed in relativistic nuclear collisions can be obtained considering
strange quark and anti-quark clusters, since they are more sensitive
to the environment from which they emerge \cite{RD87}. Recently, the
relative abundances of strange and multi-strange  baryons and 
anti-baryons where studied experimentally \cite{WA85,WA85new}.  It has
been suggested that the observed particle abundances are in agreement
with a picture of explosively disintegrating QGP fireball
\cite{Raf91}. This postulate was supported by the observation that
details of the produced particle multiplicity point to a high entropy
primordial phase \cite{Let93}, and by a comprehensive analysis of the
data which concluded that the hadron gas (HG) model cannot be brought
into consistency with the experimental results \cite{Let94}.
 
On the other hand, the alternative has been discussed 
\cite{LTR92,CS93,Cley93} that the early experimental results 
\cite{WA85}, are compatible with the simplest possible scenario of an
equilibrium, no flow, HG fireball being the particle source at which
time it was noted that it is possible to distinguish the QGP and HG
phases using the entropy content of the fireball \cite{LTR92}. We will
show here that the HG alternative is not tenable anymore, given the
larger and more precise set of experimental strange particle data
\cite{WA85new,WA85OM} and that the modifications required in order to
describe the final state hadrons point in a unique way to the
originally postulated direct (that is without re-equilibration) QGP
disintegration. We further show that the kaon multiplicity measurement
is able to fix the unknown temperature of strange particle freeze-out.
  
Implicit in the physical picture employed here (see Ref.\,\cite{Let94}
for more details) is the formation of a central and hot matter
fireball in the nuclear collisions at 200 GeV A. This reaction picture
presumes that a fraction of energy and flavor (baryon number) content
of the central rapidity region is able to thermalize; this does not
imply that the number of particles reaches ``equilibrium''
abundance characterized by the statistical variables of the system:
the temperature $T$ and the chemical fugacities $\lambda_i$ of the
different conserved quark flavors $i=u,d,s$. We recall that there is
only a slight asymmetry in the number of $u$ and $d$ quarks in the
heavy nuclei used in experiments: it is thus convenient to introduce
the quark fugacity $\lambda_{\rm q}^2=\lambda_{\rm u}\lambda_{\rm d}$
and to confine the asymmetry between the number of neutrons and
protons to the parameter $\delta\lambda\le0.03$ with $\lambda_{\rm
d}/\lambda_{\rm u}=(1+\delta\lambda)^2$ \cite{Let94}.
 
Analysis of the  S--S \cite{Sol93} and S--W data \cite{Let94} obtained
at 200 GeV A has shown that the strange quark fugacity $\lambda_{\rm
s}\simeq 1$, which is not the case for the lower energy results
\cite{RD93}. We recall that the recently reported 
$\overline{\Omega}/\Omega$ result \cite{WA85OM} has provided not only
a consistency confirmation of the thermal model but the finding
$\overline{\Omega}/\Omega=0.6\pm0.4$ which depends solely on 
$\lambda_{\rm s}$ is also independently an indication that 
$\lambda_{\rm s}\sim 1$ \cite{Let94}. The finding $\lambda_{\rm s}=1$
is of particular importance as it is natural for a directly
disintegrating QGP phase: in QGP the symmetry between the $s$ and
$\bar s$ quarks is reflected naturally in this value of $\lambda_{\rm
s}$, and only if no HG re-equilibration ensues can this value be
preserved in the final state particle abundances. This is the case
since in the HG phase, whatever the equation of state, $\lambda_{\rm
s}=1$ is an exceptional condition. At final baryon number ({\it viz.}
$\lambda_{\rm d}, \ \lambda_{\rm u}\ne 1$) the strangeness
conservation constraint requires that the number of strange and
anti-strange valance quarks bound in final state hadrons are equal.
This is in general incompatible with $\lambda_{\rm s}=1$ as is most
easily seen noting that the number of strange quarks contained in
strange baryons is not equal to the number of anti-strange quarks
contained in strange anti-baryons --- however, this asymmetry may be
compensated at some temperature $T$ by a similar asymmetry in the kaon
yields, hence non-trivial conditions may exist with $\lambda_{\rm
s}=1$ in the HG phase.
 
We now turn to the characterization of the final hadronic state in
terms of the statistical parameters. The most convenient way is to
size up the phase space in terms of all hadronic resonances, but to
allow for flexibility regarding relative particle abundances.
Throughout, we shall use the Boltzmann approximation for particle
spectra. The distribution of the final state strange particles can be 
obtained by considering the Laplace transform of the phase space
density, which leads to partition function  ${\cal Z}_{\rm s}$ as
given out of chemical equilibrium:
\begin{eqnarray}
\ln{\cal Z}_{\rm s} = { {V T^3} \over {2\pi^2} }
\left\{(\lambda_{\rm s} \lambda_{\rm q}^{-1} +
\lambda_{\rm s}^{-1} \lambda_{\rm q}) \gamma_{\rm s} C^{\rm s}_{\rm M}
F_K +(\lambda_{\rm s} \lambda_{\rm q}^{2} +
\lambda_{\rm s}^{-1} \lambda_{\rm q}^{-2}) \gamma_{\rm s} C^{\rm
s}_{\rm B} F_Y \right.\nonumber \\ 
\left.+ (\lambda_{\rm s}^2 \lambda_{\rm q} +
\lambda_{\rm s}^{-2} \lambda_{\rm q}^{-1}) \gamma_{\rm s}^2 C^{\rm
s}_{\rm B}  F_\Xi + (\lambda_{\rm s}^{3} + \lambda_{\rm s}^{-3})
\gamma_{\rm s}^3 C^{\rm s}_{\rm B} F_\Omega\right\}\ , 
\label{4a}
\end{eqnarray}
where the kaon ($K$), hyperon ($Y$), cascade ($\Xi$) and omega
($\Omega$) degrees of freedom are included successively. In the
resonance sums all known strange hadrons are counted --- in particular
we have included kaons up to 1.78~GeV, hyperons up to 1.94~GeV,
cascades up to 1.95~GeV and also the omega resonance at 2.25~GeV. The
phase space factors $F_i$ of the strange particles are explicitly
\begin{eqnarray}
F_K&=&\sum_j g_{K_j} W(m_{K_j}/T)
\ ,\qquad
F_Y=\sum_j g_{Y_j} W(m_{Y_j}/T)
\ ,\nonumber\\
F_\Xi&=&\sum_j g_{\Xi_j} W(m_{\Xi_j}/T)
\ ,\qquad
F_\Omega=\sum_j g_{\Omega_j} W(m_{\Omega_j}/T)
\ ,
\label{FSTR}
\end{eqnarray}
with $W(x)=x^2K_2(x)$, where $K_2$ is the modified Bessel function,
arising as an integral over the free Boltzmann particle phase space.
Compared to the case of an equilibrated source, several novel features
are added to allow for the possibility that the final state particles
emerge rapidly from a source of quite different nature than HG:
\begin{remunerate}
\item the factor $\gamma_{\rm s}$ in Eq.\,\ref{4a} \cite{Raf91} allows
us to consider the strange particles emerging away from absolute
chemical equilibrium which corresponds to the value $\gamma_{\rm
s}=1$. In general, if there is not sufficient time to make strangeness
(but sufficient time to exchange strange quarks between the carriers,
which we implicitly assumed above) the partition function applies with
$\gamma_{\rm s}<1$. The value of the factor $\gamma_{\rm s}$ is
determined by the dynamics of strangeness production. Its measurement
is only possible in the comparison of abundances of hadrons comprising
different numbers of strange (or anti-strange) quarks. The value
$\gamma_{\rm s}=0.7\pm0.1$ \cite{Let94} arising from the WA 85 results
\cite{WA85} is suggestive of QGP based strangeness production
mechanisms.
\item the factors $C_i^{\rm s}$ control the abundance of
strange mesons $i=$ M and baryons $i=$ B with reference to the
equilibrium based expectations \cite{Let94}. A priori, these
parameters could vary from particle to particle but intuitively we can
subsume that the processes which form mesons and baryons from some
primordial source differ substantially only between (strange) mesons
and baryons. Especially when direct hadronization of a QGP is
considered, one should not expect to see both factors to be equal to
each other or to be unity.
\end{remunerate} 
 
We require that the final state hadrons contain equal number of $s$
and $\bar s$ quarks. This can be related to the parameters present in
Eq.\,\ref{4a} using:
\begin{equation} 
0=\langle s \rangle - \langle \bar s \rangle=
\lambda_{\rm s} { \partial \over {\partial \lambda_{\rm s}}}
\ln {\cal Z}_{\rm s}\ .
\label{Eq6}
\end{equation}
This is an implicit equation relating $\lambda_{\rm s}$ with
$\lambda_{\rm q}$ for each given $T$. We stress that while we use
above a partition function to size up the phase space into
which particles are emitted, this procedure does not even tacitly
imply existence of an intermediate HG phase formed in the QGP
hadronization, and it may well be that the final state particles are
emitted directly into final asymptotic states from the QGP phase.
Recall that such a reaction picture is favored by the occurrence of
the experimental value $\lambda_{\rm s}=1$.
 
We solve the condition (\ref{Eq6}) for the  $\lambda_{\rm s}=1$ case,
and obtain:
\begin{eqnarray}
\mu_{\rm B}^0=3T{\rm cosh}^{-1}\left(R^{\rm s}_{\rm C}{F_{\rm K}\over
2F_{\rm Y}} -\gamma_{\rm s} {F_{\Xi}\over F_Y}\right)\ ,
\label{zero}
\end{eqnarray}
where we introduced the baryo-chemical potential $\mu_{\rm
B}=3T\ln\lambda_{\rm q}$ and the ratio 
\begin{equation}
R_{\rm C}^{\rm s}={C_{\rm M}^{\rm s}\over C_{\rm B}^{\rm s}}\,.
\end{equation}
We note that there is a real solution only when the argument on the
right hand side in Eq.\,\ref{zero} is greater than unity. It turns out
that this condition is a sensitive function of the temperature $T$ and
of the hadronic resonances included in Eq.\,\ref{FSTR}. For any given
hadronic spectrum used to compute the phase space factors $F_i$ there
is a maximal temperature beyond which no such solution is possible. An
important role is played by the factor $R^{\rm s}_{\rm C}$ of the
dominant first term on the r.h.s. of Eq.\,\ref{zero}. The choice
$R^{\rm s}_{\rm C}=1$ can be only justified for a source which is a
well equilibrated HG. This hypothesis is inconsistent with
$\gamma_{\rm s}=0.7$ and also challenged by the value $\lambda_{\rm
s}=1$. More natural hypothesis is that the particles are emerging from
rapidly disintegrating QGP phase. The study of the entropy content of
the source points strongly in this direction as well \cite{Let93}.
 
\begin{figure}[t]
\begin{minipage}[t]{0.475\textwidth}
\vspace{-0cm}
\caption[MUBT]{ \small 
Strangeness conservation constraint in the $\mu_{\rm B}$--$T_{\rm f}$
plane. Solid line: $\lambda_{\rm s}=1$, dashed curve: $\lambda_{\rm
s}=0.98$, dotted line: $\lambda_{\rm s}=1.08$ (all with $R_{\rm
C}^{\rm s}=1,\,\gamma_{\rm s}=0.7$). Horizontal intersecting lines
correspond to $\lambda_{\rm q}=1.48\pm0.05$. Hatched area is the
region of agreement (within 1 s.d.) of the strangeness conservation
condition with the observed values $\lambda_{\rm s}=1.03\pm0.05$ and
$\lambda_{\rm q}=1.48\pm0.05$. The experimental cross is set at the
maximal possible temperature, $T_{\rm f}=0.232\pm0.005$ GeV,
consistent with $m_\bot$-spectra. Here $R_{\rm C}^{\rm s}=1.45$ in
order to assure strangeness conservation.
\label{F1}
}
\end{minipage}\hfill
\begin{minipage}[t]{0.475\textwidth}
\vspace{- 1.6cm}
\centerline{\hspace{0.2cm}\psfig{figure=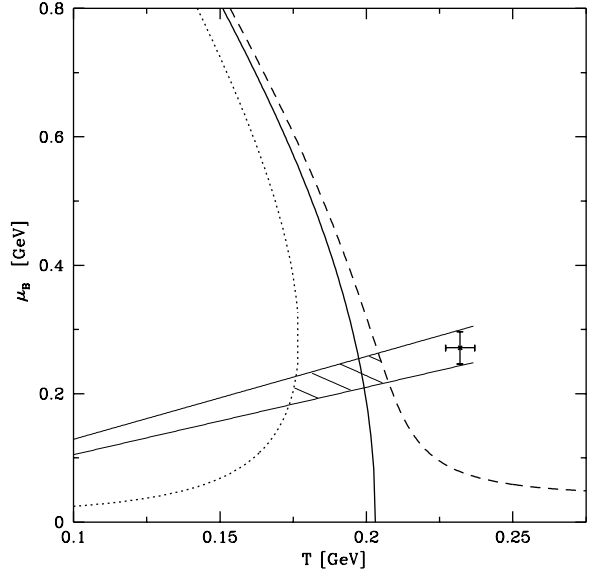,height=11.5cm}}

\vspace{ -2.cm}
\end{minipage}
\vspace{ -0.5cm}
\end{figure}

But let us first consider the conservative case and take $R_{\rm
C}^{\rm s}=1$, thus turn to the case of a HG source, and determine in
how far can the strangeness conservation constraint can be reconciled
with the full data sample of the experiment WA85
\cite{WA85,WA85new,WA85OM}. In Fig.\,\ref{F1} the solid line is just
$\mu_{\rm B}(T;R_{\rm C}^{\rm s}=1,\gamma_{\rm s}=0.7)$, see
Eq.\,\ref{zero}\,. The dashed and, respectively, dotted curves are the
solutions of the strangeness conservation condition for the
experimentally permitted range of $\lambda_{\rm s}=0.98$ and,
respectively, $=1.08$ (the mean experimental value is at $\lambda_{\rm
s}=1.03$, \cite{Let94}). The two dissecting curves correspond to the
experimental constrain $\lambda_{\rm q}=1.48\pm0.05$. The hatched area
is the region of $\mu_{\rm B}$ and $T$ compatible with a hadronic gas
source in which meson to baryon abundance equilibrium is reached. 
 
The  $m_\bot$-spectra of the strange particles used in the
multiplicity analysis indicate a common temperature $T=232\pm5$ MeV,
and this is the experimental cross, 3 s.d. to the right of the
range found to be consistent with the experiment. Thus we can conclude
that a strangeness neutral HG source (without flow) for strange
particles, even allowing for  off-equilibrium strangeness saturation
$\gamma_{\rm s}=0.7$, is in disagreement with experiment. It is at
this point interesting to note that in the first analysis of the data
\cite{WA85} the ``experimental'' point coincided with the hatched area
\cite{LTR92}.  The more precise recent data \cite{WA85new} lead to a
shift of the experimental point to the right by about 1 s.d. after a
fit of the high statistics $m_\bot$ spectra; similarly  the theory 
was also further developed and includes today all known hadronic
resonances. This has  moved  the solid line ($\lambda_{\rm s}=1$) in
Fig.\,\ref{F1} by about 1 s.d. to the left, to its current location.
Finally, the disagreement between HG-theory and experiment is enhanced
by another 1 s.d. since previously we had $\lambda_{\rm s}> 0.93$
(within 1 s.d.) and now it is $\lambda_{\rm s}> 0.98$ (dashed line in
Fig.\,1), which moves the error boundary by one s.d. to the left, to
its current location in Fig.\,1. We conclude that these recent
developments preclude the possibility of a HG interpretation of the
data, without other dynamical features, such as transverse flow
\cite{Let94}. This conclusion seems to differ from the two-temperature
hadronization model \cite{Red93}, which assumes a fully  equilibrated
(up to strangeness) sources of strange and non-strange particles at
different temperatures, and does not seem to notice that the  strange
particle $m_\bot$-spectra are too hot for the HG-model temperatures. 
 
To reach agreement today between the conditions of the source emitting
the particles and strangeness conservation constraint we must allow
for transverse surface flow. In presence of flow the thermal 
freeze-out temperature $T_{\rm f}$ of the particles would be 
blue-shifted according to the standard Doppler formula to the higher
value $T_{\rm E}$ read off the $m_\bot$-spectra:
\begin{equation}
   T_{\rm E} = T_{\rm f} \sqrt{{1+\beta_{\rm f}} 
                        \over {1-\beta_{\rm f}}}\, . 
\label{Tapp}
\end{equation}
Here $\beta_{\rm f}$ is the surface transverse velocity. The hatched
area in Fig.\,\ref{F1} is then indicating a thermal
freeze-out temperature $T_{\rm f}=190\pm 15$ MeV; such a high 
$T_{\rm f}$, though consistent with the notion of strangeness
conservation (when one subsumes HG equilibrium meson and baryon
abundances), cannot be physically correct, since it entails a much too
high particle density. Should a QGP phase hadronize at this
temperature, it must be expected that there would be continued
hadronic particle interactions and therefore collective (transverse)
flow of the resulting HG matter, with final particle decoupling
occurring at lower temperatures, where density is low. Such a scenario 
involving HG final state evolution could not be expected to result in
the $\lambda_{\rm s}= 1$ which has been observed.
  
We believe that the value $\lambda_{\rm s}\simeq 1$ may be taken as
suggesting a primordial phase hadronization in a manner in which the
emerging hadrons immediately decouple from each other. One can imagine
here two extreme cases: a freeze out at very low temperature, where
the particle density is naturally low, or a sequential hadronization
at high temperature, such that the produced hadronic particles are
escaping freely. All this can in general be accomplished in
consistency with strangeness conservation principle when allowing for
the production of strange mesons and baryons away from abundances
dictated by the relative HG chemical equilibrium. These putative
scenarios are consistent with the diverse experimental facts and
theoretical believes, but have a lot of freedom in the choice of
values for the hadronization parameters. The issue is to find a
physical observable which would allow to restrict considerably the
conditions at which strange particle hadronization occurs. This could
be clearly accomplished if the temperature of the freeze-out were to
be determined.
 
\begin{figure}[t]
\begin{minipage}[t]{0.475\textwidth}
\vspace{-1.6cm}
\centerline{\hspace{0.2cm}\psfig{figure=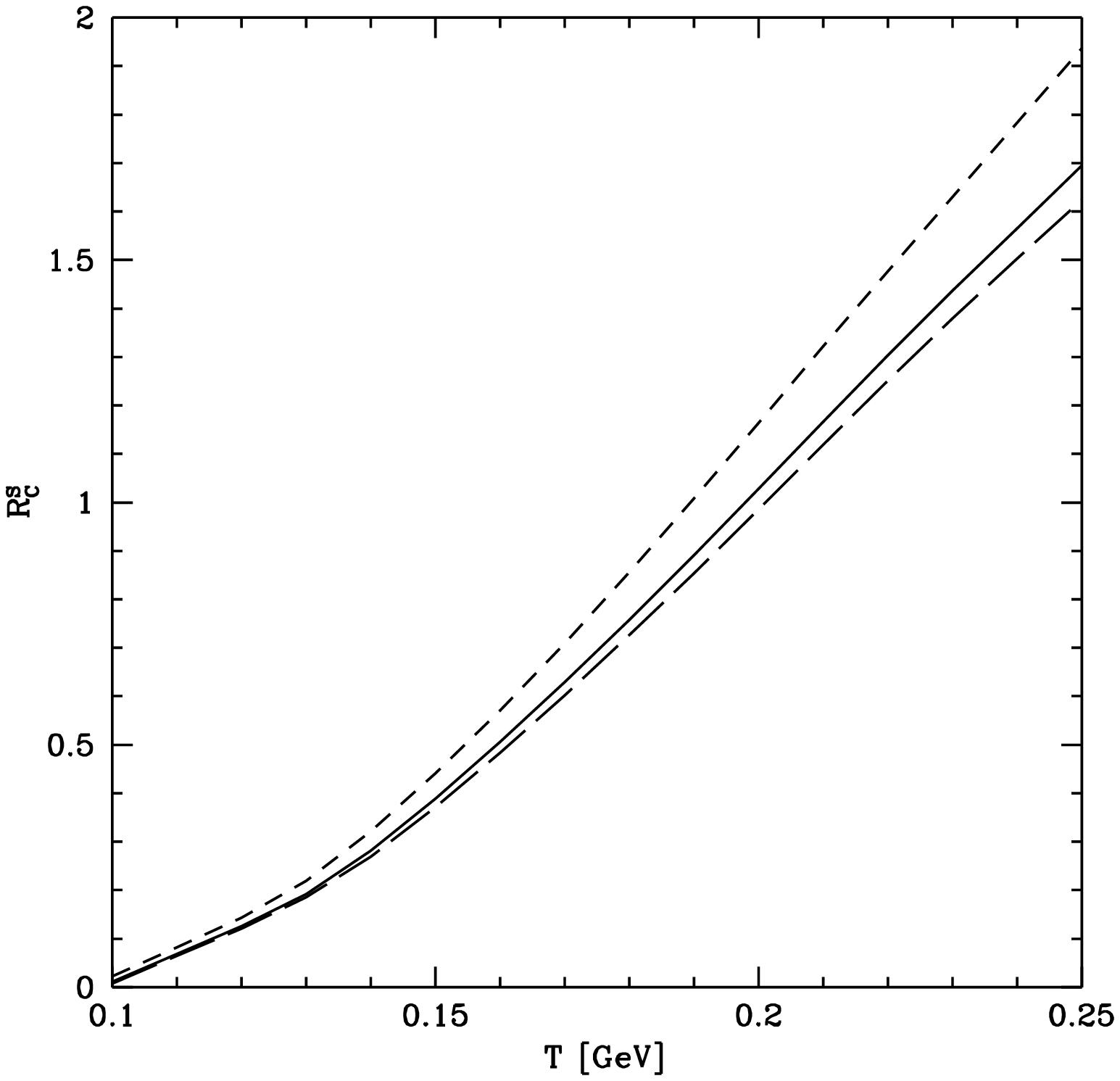,height=11.5cm}}
\vspace{ -2.cm}
\caption[RCT] 
{\small For each value of freeze-out temperature we need a different
non-equilibrium parameter $R_{\rm C}^{\rm s}$. Solid line for
$\lambda_{\rm q}=1.5$, long-dashed line for $\lambda_{\rm q}=1.3$,
short dashed curve $\lambda_{\rm q}=2$; all for $\lambda_{\rm
s}=1,\,\gamma=0.7$\,. 
\label{F2}
}
\end{minipage}\hfill
\begin{minipage}[t]{0.475\textwidth}
\vspace{- 1.6cm}
\centerline{\hspace{0.2cm}\psfig{figure=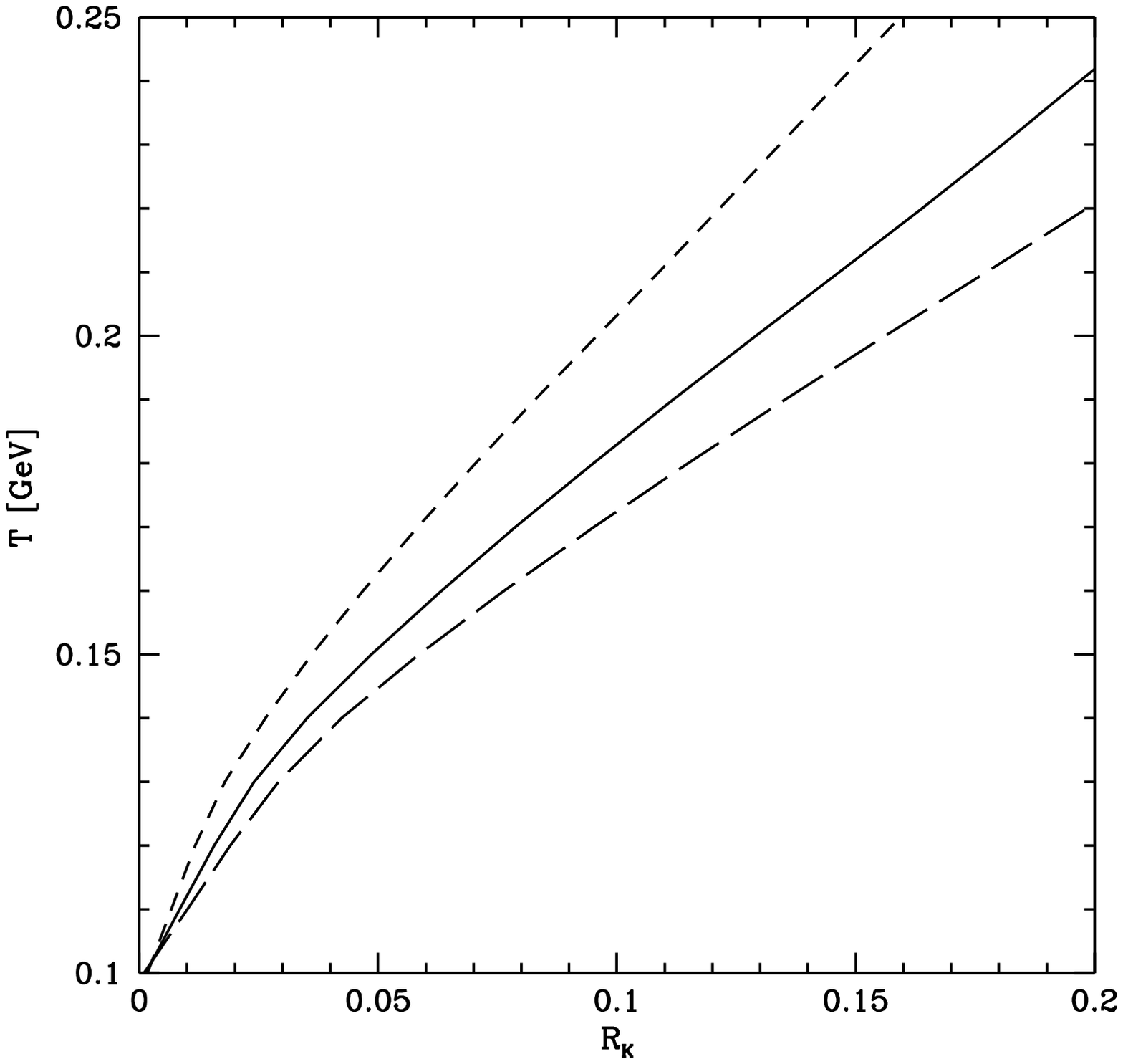,height=11.5cm}}
\vspace{ -2.cm}
\caption[TRK] 
{\small Each measured value of $R_{\rm K}$, the K-short to Lambda
ratio, implies a different value of the strange particle freeze-out
temperature. Lines as in Fig.\,\ref{F2}.
\label{F3}
}
\end{minipage}
\vspace{ -0.5cm}
\end{figure}
We first determine which values of $R_{\rm C}^{\rm s}$ are required to
have strangeness conservation at a given thermal freeze-out
temperature $T_{\rm f}$. We take the experimentally motivated
$\gamma_{\rm s}=0.7$, though the deviation from unity is of little
numerical importance in this argument. Since the (multi-)strange
(anti-)baryon particle ratios normally fix $\lambda_{\rm q}$ and
$\lambda_{\rm s}$ we take these as given: we continue to use the QGP
freeze-out value $\lambda_{\rm s}=1$ and set $\lambda_{\rm q}$ to
three values in Fig.\,\ref{F2}: the solid line is for $\lambda_{\rm
q}=1.5$, an appropriate choice for the case of S--W collisions at 200
GeV A, (when $\lambda_{\rm s}=1$), the long dashed line is for
$\lambda_{\rm q}=1.3$ which we believe is nearly appropriate for the
case of S--S collisions at 200 GeV A \cite{Sol93} (it would lead to a
central rapidity, high-$m_\bot$ particle ratio
$\overline{\Lambda}/\Lambda=0.35$). The short dashed curve is for the
choice $\lambda_{\rm q}=2$ which is our rough guess for the case of
Pb--Pb 170 GeV A collisions (and which would lead to
$\overline{\Lambda}/\Lambda=0.063$). We see that for this wide range
of physical situations ($0.35 \le \overline{\Lambda}/\Lambda \le
0.063$) we are finding rather similar  values of $R_{\rm C}^{\rm s}$
for each given freeze-out temperature. 
 
The most accessible physical observable which is sensitive
to the parameter $R_{\rm C}^{\rm s}$, and which is insensitive to
other parameters of the thermal model \cite{Let94} is the ratio 
\begin{eqnarray}
    R_{\rm K} \equiv
    {K^0_{\rm s}\over \Lambda+\Sigma^0} = {R_{\rm C}^{\rm s}
    \over 8}\,{\lambda_{\rm s}/\lambda_{\rm d} +
               \lambda_{\rm d}/\lambda_{\rm s} \over
               \lambda_{\rm s} \lambda_{\rm u} \lambda_{\rm d}} \, .   
\label{rk}
\end{eqnarray}
where the second identity is only valid if resonance decay
contributions largely cancel, which is the case \cite{Let94}. For the 
rather light kaons there exist many different possibilities for
secondary production through resonance decays, which limits somewhat
the practical usefulness of this ratio, except at very high
$m_\bot^{\rm cut}>1.9$ GeV. However, since the data of the experiment
WA85 can be constrained to such large values of $m_\bot^{\rm cut}$, we
show in Fig.\,\ref{F3}, how the freeze-out temperature $T_{\rm f}$ is
determined by the {\it measured} ratio $R_K$ (it should be here kept
in mind that along each line shown, for fixed $\lambda_{\rm
s}=1,\,\lambda_{\rm q}=\{2, 1.5, 1.3\}$ the ratio 
$\overline{\Lambda}/\Lambda$ is given as stated above, once
experimental  results become available for Pb--Pb case, a more
appropriate value of  $\lambda_{\rm q}$ may be considered).
 
The precise determination of the ratio $R_{\rm K}$ at high $m_\bot$
will in our opinion measure the hadronization temperature of the
strange particles as is indicated in Fig.\,\ref{F3}, and in this way a
full characterization of the thermal source is accomplished. For S--W
collisions, a value $R_{\rm K}\simeq 0.05$ (at $m_\bot>1.9$ GeV would
be consistent with the notion of hadronization/disintegration of the
primordial (QGP) source at temperature of about 0.15 GeV. This leads
to a value $R_{\rm C}^{\rm s}\simeq 0.4$, which implies, keeping the
meson yield near the hadronic chemical equilibrium, that at this
``low'' temperature the hadronization process enriches by a factor 
2.5 the naive thermal yield of strange baryons and anti-baryons.
Should a considerably larger values of $R_{\rm K}$ (and hence also of
$T_{\rm f}$) be found, it would suggest that the hadronization occurs
by direct and very probably sequential evaporation of particles from
the hot primordial (QGP) phase.
 
We have shown that in order to arrive at a satisfactory model of a low
temperature rapidly hadronizing primordial phase we need to consider 
off-equilibrium strange particle abundances. We have proposed a method
to measure the relevant parameter by determining precisely the central
rapidity, high $m_\bot$, K$_s$ to hyperon ratio. The sensitivity of
this measurement is sufficient to allow to fix the freeze-out
temperature of strange particles. We favor a low value at this time in
order to have a consistent explanation of the finding $\lambda_{\rm
s}\simeq 1$\, in terms of a globally hadronizing state, with resulting
particles escaping without forming an equilibrated HG phase. We await
with interest the (precise) experimental results \cite{kaon} which
will settle in our opinion the choice of hadronization model and
conditions. 
  
\vspace{0.5cm}
{\bf Acknowledgement}: J. R. acknowledges partial support by  DOE,
grant DE-FG02-92ER40733 and thanks his co-authors for their kind 
hospitality in Paris.


\vfill\eject


\end{document}